\documentclass[12pt,tightenlines,
    twocolumn,
    onecolumn,
    preprintnumbers,
    amsmath,
    amssymb,
    nofootinbib,
    prd,
    showpacs
]{revtex4}
\usepackage[dvips]{graphicx}
\usepackage{color}
\usepackage{amsmath}
\usepackage{amssymb}
\usepackage{bm}
\usepackage{enumerate}

\newcommand{\M}{{\cal M}}

\newcommand{\beq}{\begin{eqnarray}}
\newcommand{\eeq}{\end{eqnarray}}
\newcommand{\bea}{\begin{eqnarray}}
\newcommand{\eea}{\end{eqnarray}}

\begin{document}

\title{
An analytic approach to perturbations from an initially
 anisotropic universe
}
\author{Hyeong-Chan Kim}
\email{hckim@cjnu.ac.kr}
\affiliation{School of Liberal Arts and Sciences, Chungju National University,
Chungju 380-702,
  Korea}
\author{Masato Minamitsuji}
\email{masato.minamitsuji@kwansei.ac.jp}
\affiliation{Department of Physics, School of Science and Technology,
Kwansei Gakuin University, Gakuen 2-1, Sanda, 669-1337 Japan}

\date{\today}%
\bigskip
\begin{abstract}

We present the analytic forms for the spectra of the cosmological perturbations
from an initially anisotropic universe
for the high momentum modes
in the context of WKB approximations,
 as the continuation of the work \cite{km}.
We consider the Einstein gravity coupled to a light scalar field.
We then assume that the scalar field has the zero velocity initially and then slowly rolls down on the potential toward the origin.
In the slow-roll approximations, the Kasner-de Sitter universe with a planar symmetry
is a good approximation as the background evolution.
Quantization of the perturbations in the
adiabatic vacuum, which we call the anisotropic vacuum, is carried out.
For non-planar high momentum modes whose comoving momentum component
orthogonal to the plane
is bigger than the Hubble parameter at the inflationary phase,
the WKB approximation is valid for the whole stage of the isotropization.
On the other hand,
the planar modes whose comoving momentum component orthogonal to the plane
is comparable to the Hubble parameter,
is amplified during the process of the anisotropic expansion.
In the final gravitational wave spectra, we find that there is an asymmetry
between the two polarizations of the gravitational wave because
the initial mode mixing does not vanish.

\end{abstract}
\maketitle

\section{Introduction}

Recent measurements by the WMAP satellite
\cite{komatsu,wmap5,anomaly1} have suggested
that the observed map of cosmic microwave background (CMB) anisotropy
is almost consistent
with the Gaussian and statistically isotropic
primordial fluctuations from inflation.
issues on
a few anomalies
in the CMB temperature map on large angular scales
found in the recent data have been controversial.
The most well-known fact is that there seems to
be the suppression of the observed power of CMB anisotropy
on angular scales bigger than sixty degrees
\cite{anomaly1}.
There are other observational
facts that imply the effect which induces
the violation of the rotational invariance.
More precisely,
the planarity of lower multipole moments, the alignment
between the quadrupole ($\ell=2$) and the octopole ($\ell=3$),
and the alignment of them with the equinox and the ecliptic plane \cite{anomaly2} were announced.
There are other observational facts implying the large-scale anisotropy, i.e.,
odd correlations of $\ell=4\sim 8$ multipoles with $\ell=2,3$ multipoles
\cite{copi},
a very large, possibly non-Gaussian cold spot in 10 degree scale
\cite{cruz},
asymmetry of angular map measured in north and south hemispheres
\cite{eriksen}
(see also the recent review \cite{review} and
more recent references therein).
It also should be noted that
some authors claim that there is no significant evidence for primordial isotropy breaking in five-year WMAP data~\cite{Picon}
(see also more recent arguments \cite{gawe}).
Indeed, to explain the origin of the anomalies, various
solutions have been suggested,
introducing a nontrivial topology \cite{topology},
a local anisotropy based on the Bianchi type VII${}_h$ universe
to explain the quadrupole/octopole planarity and alignment \cite{jaffe},
non-linear inhomogeneities \cite{moffat}
and assuming an elliptic universe
to explain the suppression of
the quadrupole CMB power \cite{eli}.
More recently, in particular,
models which introduce an explicit source
to break
the spatial isotropy,
either during inflation or
in the late time universe,
have been proposed,
e.g.,
by the dynamics of an
anisotropic energy-momentum component during inflation
\cite{ack,anomaly,ys,wks},
by the large scale magnetic field \cite{vector},
by the anisotropic cosmological constant \cite{cc}
or dark energy \cite{de}.

The first purpose of this paper is to proceed to investigate
the possibility that such large scale anomalies
are produced by
preinflationary anisotropy
and obtain the leading order corrections
to the spectra for them. 
Cosmic nohair theorem ensures
that in the presence of a positive cosmological constant
an initially anisotropic universe exponentially
approaches the de Sitter spacetime at the later time
under the strong or dominant energy condition \cite{wald}.
Therefore, it is plausible that
the initial universe is highly anisotropic.

The future CMB measurements will
detect the fluctuations of B mode polarization
in CMB, which may contain
the information on the primordial gravitational waves.
They would give a new tool to contrain
the anisotropic cosmological model.
The cosmological perturbation theory in the
Kasner phase was formulated in Ref. \cite{tpu,gcp,gkp}.
In general in an expanding (planar) Kasner phase
one of two polarizations of gravitational waves
is coupled with the scalar mode,
but the other gravitational mode is decoupled.
Thus, this coupling induces
the asymmetry between propagations of two polarizations of the
gravitational waves.
If there are
effects of the chiral symmetry breaking in the cosmic history,
they would give rise to nonzero cross correlations
between the fluctuations
of the temperature and B mode,
and
of E and B modes \cite{lue}.
They will give us powerful and
independent tests on the primordial parity violation.
The gravitational waves from the universe with an isotropy breaking
would provide
distinguishable signatures in the future CMB
experiments.
We will estimate how the initial mode mixing
gives rise to the asymmetry between the
primordial power spectra of the two gravitational wave modes,
although we will not go into details of
the observational aspects.
Note that
such subjects
have been argued in the context of
the anisotropic inflation models in Ref. \cite{aniso_inf}.
The investigation of
the higher order correlations as the bispectrum
would provide
us another interesting prediction to examine the
the anisotropic universe
(see e.g., \cite{chen}).
We briefly comment on some expectations on this point in the last section.

In the isotropic case, the quantization of fluctuations is carried out
well inside the Hubble horizon,
where the effects of the cosmic expansion can be ignored.
In order to compare with the standard prediction,
it is natural to quantize field in the initial adiabatic vacuum,
which we call the {\it anisotropic vacuum}.
There are two branches of the expanding Kasner solution
with the planar symmetry.
The initial adiabatic vacuum present only on one of the two branches where
the expansion rate along the planar directions vanishes while that along the special axis is finite
\footnote{
The anisotropic vacuum is not specific to the Bianchi I model.
In Ref. \cite{km},
it has been shown that
an anisotropic vacuum
can also be defined for a Bianchi IX model.}.
As a result, the initial spacetime structure can be seen as the product of two-dimensional Milne spacetime and two-dimensional Euclidean space.
The scalar fluctuations decouple from the tensor fluctuations
at the very initial time, so that
the initial dynamics reduces to that in
a system composed of three independent harmonic oscillators.
In the other branch there is an initial singularity.
Since the coupling diverges at the initial time,
we cannot find an adiabatic vacuum.
Therefore,  here we focus on the first branch.
For a given set of initial conditions, the power spectrum was investigated,
rather by the numerical ways in Ref. \cite{tpu,gcp,gkp}.
The aim of our study is to obtain more analytic understandings
on the spectra from an initially anisotropic universe.
Our previous work discussed the spectrum of a massless scalar field,
ignoring its coupling with the metric perturbations \cite{km}.
In this work, as the continuation, we will discuss the metric perturbations,
in particular focusing on the importance of the tensor-scalar coupling.

The paper is constructed as follows:
In Sec. II, the background solution of our anisotropic model
is introduced.
In Sec. III,
we present the formulation of the coupled perturbations
in the background of Kasner de Sitter solution with the planar symmetries
and their relation to the cosmic observables.
In Sec. IV, we investigate the behaviors of the perturbations modes
after setting initial conditions in the anisotropic vacuum.
In Sec. V, we close the article after giving a brief summary.

\section{Background}

We consider the Einstein gravity
minimally coupled to a massive scalar field
\bea
S=\int d^4x \sqrt{-g}
\Big(\frac{M_p^2}{2}R-\frac{1}{2}(\partial\phi)^2
-\frac{1}{2}m^2\phi^2
\Big)\,,
\eea
where $g_{\mu\nu}$ is the spacetime metric and
$\phi$ is a canonical scalar field with mass $m$.
We consider an anisotropic spacetime with a two-dimensional
planar symmetry
\bea
\label{metric}
ds^2=-d\tau^2+a(\tau)^2 dx^2+b(\tau)^2 (dy^2+dz^2),
\eea
where $a(\tau)$ and $b(\tau)$ are
independent scale factors.
We define the expansion rates by
\bea
H_a:=\frac{\dot{a}}{a},\quad
H_b:=\frac{\dot{b}}{b},
\label{rel}
\eea
respectively.
In this paper, ``dot'' denotes
the derivative with respect to the proper time $\tau$.
Then, the field equations are given by
\bea
&&{\dot H}+3H^2=\frac{m^2\phi^2}{2M_p^2},
\quad
3H^2-h^2=\frac{1}{M_p^2}
\Big(\frac{1}{2}\dot{\phi}^2+
\frac{m^2\phi^2}{2}\Big),
\quad
\ddot{\phi}+3H\dot{\phi}
+m^2\phi
=0,
\label{ora}
\eea
where
the total and relative expansion rates are defined as
\bea
\label{rel2}
H
:=\frac{H_a+2H_b}{3},\quad
h:=\frac{H_a-H_b}{\sqrt{3}}.
\eea
Note that the evolution of the scalar field does not depend on the anisotropic scales but depends on the averaged scale factor.

We assume that the scalar field $\phi$
stays initially at $\phi=\phi_0>M_p$,
and then slowly rolls down.
Under this approximation, we can ignore the kinetic energy
of the scalar field in the second equation of \eqref{ora},
and $\phi\approx \phi_0$.
From the first and then
second equations of \eqref{ora},
we obtain
\bea
H
=H_0 \coth (3H_0 \tau),
\quad
h
=\pm\frac{\sqrt{3}H_0}{\sinh(3H_0\tau)},
\eea
where
\bea
H_0=\frac{m\phi_0}{\sqrt{6}M_p},
\label{late_h}
\eea
is the total
Hubble parameter at the asymptotic region
in the limit of $\tau \to \infty$.
Here, (-)-branch leads to the solution
which contains a curvature singularity at the initial time
which will not be dealt with in this paper.
For the (+)-branch, we obtain
\bea
&&H_a= \frac{H_0}{\sinh(3H_0 \tau)} \Big( 2+\cosh(3 H_0 \tau) \Big),
\quad
H_b = H_0 \tanh\big(\frac{3}{2}H_0 \tau\big).
\eea
At the initial time, they reduce to
$H_a\to \frac{1}{\tau}$ and $H_b\to 0$,
which represents a (Milne) patch of the Minkowski spacetime
and does not contain an initial singularity.
The two-independent scale factors are given by
\bea \label{yoko2}
a=\sinh ^{1/3}(3H_0 \tau)
 \tanh^{\frac{2}{3}} \Big(\frac32 H_0 \tau\Big),\quad
b= \sinh ^{1/3}(3H_0\tau) \coth^{\frac{1}{3}}
\Big(\frac{3}{2}H_0 \tau\Big).
\eea
The averaged scale factor is given by
\bea
e^{\alpha}:=(ab^2)^{\frac{1}{3}}
=\big(\sinh (3H_0\tau)\big)^{\frac{1}{3}}.
\eea

On this background, the evolution of the scalar field
is approximated well by the behavior of
a massive field in the above background
\bea
\phi&=&
\frac{\phi_0}{\sqrt{\pi}\tan\big(\frac{\pi q}{2}\big)}
\Big(
\frac{\Gamma\big(\frac{1-q}{2}\big)}
     {\Gamma\big(1-\frac{q}{2}\big)}
e^{\frac{1}{4}(1-q)x}
{}_2F_1\big[\frac{1}{2},\frac{1-q}{2},1-\frac{q}{2},e^x\big]
\nonumber\\
&-&
\frac{\Gamma\big(\frac{1+q}{2}\big)}
     {\Gamma\big(1+\frac{q}{2}\big)}
e^{\frac{1}{4}(1+q)x}
{}_2F_1\big[\frac{1}{2},\frac{1+q}{2},1+\frac{q}{2},e^x\big]
\Big),\label{phi1}
\eea
where we defined
$x:=6H_0\tau$,
$q:=\sqrt{1-16M^2}$
and
$M:=\frac{M_p}{\sqrt{6}\phi_0}
=\frac{m}{6H_0}$.
For $q$ to be real,
$\phi_0>\sqrt{\frac{8}{3}}M_p$ {\it i.e.,}
$m< \frac{3H_0}{2}$.
In the anisotropic stage
Eq. (\ref{phi1}) can be further approximated as
\bea
\phi\approx
\phi_0\big(1-\frac{1}{4}m^2\tau^2\big).
\label{phi2}
\eea
At the late time after the isotropization,
$\tau\gg 1/H_0$ ($x\gg 1$),
$
\phi\propto e^{-\frac32 H_0 \tau
\big(1-\sqrt{1-16M^2}\big)}$.
As long as
\begin{eqnarray}\label{SlowRoll}
\phi_0\gg\sqrt{\frac{8}{3}}M_p,
\end{eqnarray}
namely, $m\ll\frac{3H_0}{2}$,
the slow-roll condition is satisfied.

Let us check the consistency of
our approximation.
The typical time scale for the cosmic isotropization is
given by $x_{\rm iso}=1$,
namely $\tau_{\rm iso}=\frac{1}{H_0}$.
On the other hand,
the kinetic energy of the scalar field can
be comparable to the potential
at the time scale $\tau_{\phi}=\frac{1}{m}$.
Therefore, the condition that
$\phi$ almost stays around $\phi_0$
and the kinetic energy of the scalar field is negligible
is given by $\tau_{\phi}\gg \tau_{\rm iso}$,
namely
$\phi_0\gg\frac{M_p}{\sqrt{6}}$ from Eq. (\ref{late_h}).
Thus, it turns out that
with the slow-roll condition during inflation Eq. (\ref{SlowRoll}),
under the assumption of the super-Planck initial amplitude,
the solution Eq. (\ref{yoko2})
becomes a good approximation
for the background evolution of the
preinflationary universe.

\section{Perturbations}


\subsection{Perturbations in the Bianchi-type I universe}

\subsubsection{Mode decompositions}

We then consider the cosmological perturbation theory in the
anisotropic spacetime with a planar symmetry,
following Ref. \cite{gcp}.
The decomposition into the independent modes is performed
in terms of the two-dimensional $(y,z)$-plane.
A scalar quantity contains 1 degree of freedom.
A vector quantity which satisfies the transverse condition
in two dimensions
also contains 1 degree of freedom.
But there is no degree of freedom for a tensor quantity
satisfying the transverse-traceless condition in two dimensions.
Under the above decomposition,
the totally 10 components of the metric perturbations
can be classified into 7 scalar and 3 vector modes.
Then,
the perturbed metric can be described by
\bea
g_{\mu\nu}
=\left(\begin{tabular}{cccc}
$-a^2(1+2\Phi)$ & $a\partial_1 \chi$ & $a\partial_2 B$ & $b^2 B_3$\\
$$ & $a^2(1-2\Psi)$&$b^2\partial_1\partial_2{\tilde B}$
&$b^2\partial_1{\tilde B}_3$
\\
$$ & $$ & $b^2 (1-2\Sigma+2\partial_2^2 E_3)$ & $b^2 \partial_2 E_3$ \\
$$ & $$ & $$ & $b^2(1-2\Sigma)$ \\
                   \end{tabular}
\right),
\eea
where the matric is symmetric.
$E_3$, $B_3$ and ${\tilde B}_3$ correspond to vector modes,
and the rest are scalar ones.
In addition,
there is
the perturbation of the scalar field, $\phi+\delta\phi$,
which is definitively the scalar mode.
Thus, there are totally
8 scalar and 3 vector modes.

In the rest of the paper,
we work in the momentum space
after decomposing perturbations into (comoving) Fourier modes.
We distinguish the comoving momenta $k_i$
from physical momenta $p_i$, which depends on time, as
\bea
p_1:=\frac{k_1}{a},\quad
p_2:=\frac{k_2}{b}.
\eea
The total momenta are defined by
\bea
k^2=k_1^2+k_2^2,\quad
p^2:=p_1^2+p_2^2.
\eea
Thanks to the residual planar symmetry on the $(y,z)$ plane,
we can fix $k_3 =0$ without any loss of generality.
$k_1$ denotes the component of the momentum along the
special $x$ direction, and
$k_2$ does that in the orthogonal plane.

\subsubsection{Vector mode}

About the vector mode,
1 of 3 components can be eliminated by the gauge fixing
and the other 1 can be done by the momentum constraint.
By setting the gauge of vector mode to be $E_3=0$
and eliminating the nondynamical component $B_3$,
we have only one propagating normalized degree of freedom
\bea
H_{\times}:=\frac{M_p}{\sqrt{2}}
\frac{k_1 k_2}{\sqrt{k_1^2+\frac{a^2}{b^2}k_2^2}}
{\tilde B}_3,
\eea
whose equation of motion is given by
\bea
\Big[\frac{d^2}{dt^2}
+\omega_{\times}^2\Big] H_{\times}
=0,
\eea
with
\bea
\omega_{\times}^2
:=a^2b^4\Big[p_1^2+p_2^2-H_a(H_a-H_b) +\dot{H}_b
  +\frac{\dot{\phi}^2}{2M_p^2}+\big(H_a-H_b\big)^2
  \frac{p_1^2\big(p_1^2+4p_2^2\big)}
     {\big(p_1^2+p_2^2\big)^2}
\Big]. \label{omega cross}
\eea
For convenience,
we introduced a new time coordinate $t$ by
\begin{eqnarray}
dt=\frac{d\tau}{ab^2}
=\frac{d\tau}{e^{3\alpha}}
\end{eqnarray}
in which
$$
ab^2=\sinh\Big(
                3H_0\tau\Big)
=\Big[\sinh\Big(3H_0(-t)\Big)\Big]^{-1}.
$$
Note that $t\to -\infty$ as $\tau\to 0$, and
$t\to 0-$ as $\tau\to \infty$.

%

\subsubsection{Scalar mode}

About the scalar mode,
3 of totally 8 components can be eliminated by the gauge fixing
and the other 3 can be done by the constraints
(1 Hamiltonian and 2 momentum constraints)
and hence there are two propagating degrees of freedom.
Setting $\tilde B=\Sigma= E=0$ by fixing the gauge,
and then eliminatng the nondynamical components
$\Phi$, $\chi$ and $B$,
two normalized propagating degrees of freedom are
given by
\bea
V=
\delta \phi +\frac{p_2^2 \dot \phi}
{H_a p_2^2 +H_b(2p_1^2+p_2^2)} \Psi,
\quad
H_+ = \frac{\sqrt{2}   M_p p_2^2 H_b}
{H_a p_2^2+H_b(2p_1^2+p_2^2)}
\Psi.
\eea
$V$ and $H_{+}$ obey the coupled equations of motion
\begin{eqnarray}
\label{mat}
\left[
{\bf 1}
\frac{d^2}{dt^2}
+                \left(\begin{tabular}{cc}
                     $\omega_{11}^2$& $ \omega_{12}^2$ \\
                     $\omega_{12}^2$ & $\omega_{22}^2$\\
                   \end{tabular}\right)\right]
\left(\begin{tabular}{c}
                     $V$ \\
                     $H_+$ \\
                   \end{tabular}\right) =0,
\end{eqnarray}
where
the components of the frequency matrix are given by
\bea
\label{freq}
\omega_{11}^2
&:=&a^2 b^4\Big\{p_1^2
+p_2^2
+H_b(H_b-H_a) +\dot{H}_b
+\frac{3\dot{\phi}^2}{2M_p^2}
+\frac{2H_a}{H_b}\frac{\dot{\phi}^2}{M_p^2}
-\frac{1}{2H_b^2}\frac{\dot{\phi}^4}{M_p^4}
+\frac{2}{H_b}\frac{m^2\dot{\phi}\phi}{M_p^2}
+m^2
\nonumber\\
&+&\frac{p_2^2(H_a-H_b)}
        {2H_b p_1^2+(H_a+H_b)p_2^2}
\frac{\dot{\phi}}{M_p}
\Big[
-\frac{4\dot{\phi}}{M_p}
-\frac{2H_a}{H_b}\frac{\dot{\phi}}{M_p}
+\frac{\dot{\phi}^3}{H_b^2 M_p^3}
-\frac{2m^2\phi}{H_b M_p}
\nonumber\\
&-&\frac{p_2^2(H_a-H_b)}
        {2H_b p_1^2+(H_a+H_b)p_2^2}
\frac{\dot{\phi}}{M_p}
\Big(1+\frac{\dot{\phi}^2}{2H_b^2 M_p^2}\Big)
\Big]
\Big\},
\nonumber\\
\omega_{22}^2
&:=&
a^2 b^4
\Big\{
p_1^2+p_2^2+H_b(H_b-H_a) +\dot{H}_b
+\frac{\dot{\phi}^2}{2M_p^2}
\nonumber\\
&+&
\frac{p_2^2\big(H_a-H_b\big)^2}
      {2H_b^2 p_1^2+(H_a+H_b)p_2^2}
\Big[
 4H_b
-\frac{p_2^2\big(2H_b^2+\frac{\dot{\phi}^2}{M_p^2}\big)}
      {2H_b p_1^2+(H_a+H_b)p_2^2}\Big]
\Big\},
\nonumber\\
\omega_{12}^2
&:=&a^2b^4
\frac{\sqrt{2}p_2^2 (H_a-H_b)}
      {2H_b p_1^2+(H_a+H_b)p_2^2}
\Big[
-3\frac{H_b\dot{\phi}}{M_p}
+\frac{1}{2H_b}\frac{\dot{\phi}^3}{M_p^3}
-\frac{m^2\phi}{M_p}
\nonumber\\
&-&
\frac{p_2^2(H_a-H_b)}
        {2H_b p_1^2+(H_a+H_b)p_2^2}
\frac{\dot{\phi}}{M_p}
\Big(H_b+\frac{1}{2H_b}\frac{\dot{\phi}^2}{M_p^2}\Big)
\Big].
\eea

\subsubsection{The isotropic limit}

Before closing this section,
we briefly mention the limit to
the ordinary homogeneous and isotropic universe
$b\to a$.
In the later times,
we find
\bea
\label{latefreq}
&&\omega_{11}^2\to
a^4\Big(k^2
+\frac{m^2\phi\dot{\phi}}{M_p^2 H}
+\frac{7\dot{\phi}^2}{2M_p^2}
-\frac{\dot{\phi}^4}{2M_p^4H^2}\Big)
=a^4(k^2-\frac{z''}{z}+2H_0^2 a^2),
\nonumber\\
&&
\omega_{22}^2,\quad
\omega_{\times}^2
\to
a^4\Big(k^2
+\frac{a^2\dot{\phi}^2}{2M_p^2}\Big)
=
a^4\Big(k^2-\frac{a''}{a}+2H_0^2 a^2\Big),
\eea
where $z=\frac{a^2\dot{\phi}}{\dot{a}}$,
while $\omega_{12}\to 0$.
The prime denotes the derivative with respect to
the conformal time $d\eta=\frac{d\tau}{e^{\alpha}}$.
The late time evolutions are given in terms of those in the de Sitter spacetime,
written in terms of the Bessel functions.
Thus,
in this limit, 
$V$ reduces to the Sasaki-Mukhanov variable $v$
of scalar perturbations in three dimensions,
defined in \cite{ms},
while $H_\times$ and $H_{+}$
give two independent tensor polarizations
$h_\times$ and $h_+$ in the flat three-dimensional space.
Therefore, through the propagations in the
anisotropic universe,
an asymmetry between the two tensor polarizations
would appear.

\subsection{Quantization and power spectra}

After giving the equations of motion,
we are going to quantize the perturbation modes.
At the initial times, the frequency squares of the vector and scalar modes in
Eqs. (\ref{omega cross}) and (\ref{freq}), behave as
\bea
\label{ome}
\omega_{\times}^2 &=&2^{\frac{4}{3}}k_1^2 +
   \left(2^{\frac{1}{3}}(4k_1^2 +3k_2^2)
  +\frac{27H_0^2 k_2^2}{k_1^2}
   \right)\frac{3H_0^2\tau^2}{2}+O(\tau^3),
\nonumber\\
\omega_{11}^2 &=&2^{\frac{4}{3}}k_1^2+
  \left(2^{1/3}(4k_1^2+3k_2^2)-6 m^2\frac{4k_1^2-3k_2^2}{4k_1^2+3k_2^2}
  \right)\frac{3H_0^2 \tau^2}{2}+O(\tau^3),
\nonumber\\
\omega_{22}^2 &=&2^{\frac{4}{3}}k_1^2
+ \left(  2^{\frac{1}{3}}(4k_1^2+3k_2^2) +
 \frac{108 H_0^2 k_2^2}
{4k_1^2+3k_2^2}
 \right)\frac{3H_0^2\tau^2}{2}
+O(\tau^3),
\nonumber\\
 \omega_{12}^2
&=&-\frac{36\sqrt{3}H_0 m \,k_2^2 }
       {4 k_1^2+3 k_2^2} \frac{3H_0^2\tau^2}{2}
+O(\tau^3).
\eea
The frequency squared $\omega_\times^2$ appears to be divergent
in the limit $k_1\to 0$.
However, it is 
an artifact of the early time limit as one can see in Eq.~(\ref{omega cross}).
Since all of $\omega_{\times}^2$, $\omega_{11}^2$ and $\omega_{22}^2$
approach constants,
and $\omega_{12}^2\to 0$ as $\tau\to 0$,
namely,
the coupling between $V$ and $H_+$ vanishes in the early time, 
the adiabatic vacuum can be found.
To distinguish this vacuum state from the
standard Bunch-Davis vacuum,
we call our adiabatic vacuum
{\it an anisotropic vacuum}.
Then,
these perturbations are quantized in this vacuum.
The procedure of the canonical quantization 
follows the standard manner.

Here, we introduce $Y$ for the collective notation of
$H_\times$, $V$ and $H_+$.
For the anisotropic vacuum $|0\rangle$,
the anihilation operator can be defined by
$a_{\bf k}|0\big \rangle=0$.
Then, the metric perturbations
are canonically quantized as
\beq
Y= \int d^3k
    \Big(u_{\bf k} a_{\bf k}
+    u^{\ast}_{\bf k} a^{\dagger}_{\bf k}
\Big),
\eeq
where operators satisfy the
commuation relation $[a_{\bf k_1},a^{\dagger}_{\bf k_2}]
=\delta({\bf k_1}-{\bf k_2})$ (others are zero)
and $u_{\bf k}= e^{i{\bf k}{\bf x}}Y_{\bf k}/(2\pi)^{3/2}$
are the mode functions satisfing
the normalization condition
$
Y_{\bf k}\partial_t Y^{\ast}_{\bf k}
-\big(\partial_t Y_{\bf k} \big)Y^{\ast}_{\bf k}
=i/e^{3\alpha}.
$
Frow now on, we omit the subscript $\bf k$.
Since from Eq. (\ref{latefreq}) the mode mixing
is also absent in the late time universe,
the final power spectra of these modes
are independently defined by
\bea
P_{v}
=\Big(\frac{H_0}{\dot{\phi}}\Big)^2
\frac{b^3p^3}{2\pi^2}|V|^2\Big|_{t\to 0-},
\quad
P_{h_\times}
=\frac{b^3p^3}{\pi^2 M_p^2}|H_{\times}|^2\Big|_{t\to 0-},
\quad
P_{h_{+}}
=\frac{b^2 p^3}{\pi^2 M_p^2}
 | H_{+}|^2\Big|_{t\to 0-},
\eea
and related to the late time behaviors
of perturbations. 
For comparison,
in the standard slow-roll inflation,
the spectra are given by
\bea
\label{zeroth}
P_{v}^{(0)}=\Big(\frac{H^2}{2\pi\dot{\phi}}\Big)^2,\quad
P_{h_{+}}^{(0)}=P_{h_{\times}}^{(0)}
=\frac{H^2}{2\pi^2 M_p^2}.
\eea

\subsection{
Evolutions of perturbations in the intermediate times and WKB approximation}

In the intermediate times,
two modes are generically coupled.
If the WKB approximation is valid, however,
we may simplify to solve these perturbations.
The original equation Eq.~(\ref{mat}) can be
described by the matrix differential equation
\begin{eqnarray}\label{eom}
\frac{d^2M}{dt^2} + \Omega^2 M =0,
\end{eqnarray}
where
\bea
M=\left(\begin{tabular}{c}
                     ${ V}$ \\
                     ${ H}_+$ \\
                   \end{tabular}\right),
\quad
\Omega^2=
\left(\begin{tabular}{cc}
                     $\omega_{11}^2$& $ \omega_{12}^2$ \\
                     $\omega_{12}^2$ & $\omega_{22}^2$\\
                   \end{tabular}\right).
\eea
Defining $M = A{\cal M}$,
where
$A$ is a rotational matrix
$$
A =
\left(\begin{tabular}{cc}
                     $\cos \theta$& $ -\sin \theta$ \\
                     $\sin \theta$ & $\cos \theta$ \\
                   \end{tabular}\right) ,
$$
Eq. (\ref{eom}) becomes
\bea
 \label{ddM2}
\frac{d^2\M}{dt^2} +2A^{-1}\frac{dA}{dt}\frac{d \M}{dt}+
  \Big(A^{-1}\frac{d^2A}{dt^2}+{\tilde \Omega}{}^2\Big)\M=0.
\eea
By choosing $\theta$ so that $\tilde{\Omega}{}^2
\equiv A^{-1} \Omega^2 A$ becomes a diagonal matrix,
we obtain
\begin{eqnarray}
\tan (2\theta)=-\chi,\quad
\tan \theta =\frac1{\chi} \pm \sqrt{\frac1{\chi^2}+1},
\quad \chi:=\frac{2\omega_{12}^2}{\omega_{22}^2-\omega_{11}^2}.
\end{eqnarray}
\bea
\cos\theta
&=&\Big[\frac{1}{2}\Big(1+\frac{1}{\sqrt{1+\chi^2}}\Big)\Big]^{\frac{1}{2}},
\quad
\sin\theta
=
\pm
 \Big[\frac{1}{2}\Big(1-\frac{1}{\sqrt{1+\chi^2}}\Big)\Big]^{\frac{1}{2}}, \label{theta}
\eea
where
the positive (negative) sign will be taken for $\omega_{22}^2>\omega_{11}^2$ ($ \omega_{22}^2< \omega_{11}^2$) to have $\theta=0$ in the limit
$\chi\to \mp 0$.
Note that $\chi<0$ corresponds to $\theta>0$
and vise versa.

At the initial period of time,
\begin{eqnarray}
\chi = -\frac{2m/H_0}{\sqrt{3}\left[1+\frac{m^2}{6H_0^2} \big(\frac43 \frac{k_1^2}{k_2^2}-1 \big)\right]} +O(\tau^2)
\label{chio}
\end{eqnarray}
is of $O(m/H_0)$.
Thus, interestingly,
although $\omega_{12}\to 0$ as $t\to-\infty$,
$\chi$ remains finite in the same limit
and there is a finite amount of the mode mixing initially.
On the other hand,
for $t\to \infty$, $\chi\to 0$.
The effect of the mode mixing
is encoded into terms which contain
the time derivatives of the rotational matrix $A$ in Eq.~(\ref{ddM2}).
We also find
$$
{\tilde \Omega}{}^2=A^{-1}\Omega^2 A
=   \left(\begin{tabular}{cc}
                 $\tilde{\omega}{}_{11}^2$& $0$ \\
                 $0$ & $\tilde{\omega}{}_{22}^2$\\
                   \end{tabular}\right),
$$
where
\bea
{\tilde \omega}{}_{11}^2
&=&\frac{1}{2}\left(\omega_{11}^2+\omega_{22}^2
-(\omega_{22}^2-\omega_{11}^2) 
 \sqrt{1+\frac{4\omega_{12}^4}{(\omega_{11}^2-\omega_{22}^2)^2}}
\right),
\nonumber\\
{\tilde \omega}{}_{22}^2
&=&\frac{1}{2}\left(\omega_{11}^2+\omega_{22}^2
+(\omega_{22}^2-\omega_{11}^2)
 \sqrt{1+\frac{4\omega_{12}^4}{(\omega_{11}^2-\omega_{22}^2)^2}}
\right).
\eea
Then,
\bea
\M=\left(\begin{tabular}{c}
                     ${\tilde V}$ \\
                     ${\tilde H}_+$ \\
                   \end{tabular}\right)
=\left(\begin{tabular}{c}
                     $V\cos\theta+ H_+\sin\theta$ \\
                     $H_+\cos\theta-V \sin\theta $ \\
                   \end{tabular}\right)\,,
\eea
where
$\theta$ is a slowly varying function
as $\theta'\sim O(\omega_{11}',\omega_{22}')$
so that the WKB approximation is valid.
Thus,
\bea
&&
\Big|\frac{d^2}{dt^2}\M
\Big|\sim  |\omega^2\M|\gg
\Big|A^{-1}
\Big(\frac{d}{dt}A\Big)
\frac{d}{dt}\M
\Big|\sim
\Big|\frac{d}{dt}\omega
\M
\Big|,
\nonumber\\
&&
|A^{-1}\Omega^2 A|\sim |\omega^2\M|
\gg
\Big|A^{-1}\Big(\frac{d^2}{dt^2}A\Big)
\M| \sim
\Big|
\frac{\frac{d^2}{dt^2}\omega
}{\omega}\M
\Big|.
\eea
Therefore,
to obtain the leading order correction,
it is enough to solve
\begin{equation} \label{ddM}
\frac{d^2\M}{dt^2}
+{\tilde \Omega}{}^2
\M \approx0.
\end{equation}
If the WKB is not valid,
one has to solve Eq. (\ref{eom}) directly.

Introducing a new corrective notation $\tilde Y$
for 
$H_{\times}$, $\tilde V$ and ${\tilde H}_+$,
then we can check the validity of the WKB approximation
through 
the adiabaticity parameter,
\bea
\epsilon_{\tilde Y}
:=\frac{|\frac{d}{dt}\omega_{\tilde Y}^2|}{(\omega_{\tilde Y}^2)^{\frac{3}{2}}},\label{ups}
\eea
where $\omega_{\tilde Y}^2$
represents
either of $\omega_{\times}^2$,
${\tilde \omega}_{11}^2$ or ${\tilde \omega}_{22}^2$.
For the high momentum case,
since the behaviors of $\omega_{\tilde Y}$ are very similar to
to the case of a massless scalar field discussed in~\cite{km}
from Eq.~(\ref{ome}),
$\epsilon_{\tilde Y}$ also behaves as in the same way (and see Fig. I).
Hence, for the {\it non-planar} high-momentum modes $k_1\sim k_2\gg H_0$,
$\epsilon_{\tilde Y}$ is always less than unity
in the early times,
while for the {\it planar} modes $H_0<k_1\ll k_2$
it temporarily exceeds unity,
which implies that the WKB approximation is broken there.
Therefore, in the next sections
we treat these two cases separately.


\section{Non-planar modes with $k_1, k_2 \gg H_0$}

For the non-planar high momentum mode, the WKB approximation is valid
for a small scale factor.
We can discuss the evolution of the perturbation
modes by solving Eq.~(\ref{ddM}).
Since the equations are approximately diagonalized,
we solve them correctively.

In the early time, all $H_{\times}$, $V$ and $H_{+}$ modes are quantized
in the adiabatic vacuum.
Their initial amplitudes are given by
\bea
{H}_{\times}\big|_{t\to -\infty}
=
{V}\big|_{t\to -\infty}
=
H_{+}\big|_{t\to -\infty}
=\frac{1}{\sqrt{2k_1}}.
\eea
On the scalar mode from Eq. (\ref{chio}),
at the initial time
the mixing angle is not vanishing, $\sin\theta \approx 
\frac{m}{\sqrt{3}H_0}$
and $\cos\theta\approx 1$,
hence initial amplitudes of modes
after the diagonalization are given by
\bea
{\tilde V}\big|_{t\to -\infty}
\approx \Big(V
+\frac{m}{\sqrt{3}H_0}H_{+}\Big)\big|_{t\to -\infty},
\quad
{\tilde H}_+\big|_{t\to -\infty}
\approx \Big(H_+
-\frac{m}{\sqrt{3}H_0}V\Big)\big|_{t\to -\infty},
\eea
respectively.
For the high momentum modes, ignoring the corrections of $O(m^2)$,
$\omega_{11}= \omega_{22}={\tilde \omega}_{11}= {\tilde\omega}_{22}$
which are correctively represented by $\omega$.
In addition, since the WKB approximation is valid for the
non-planar high momentum modes, we obtain
the time evolution in the intermediate times as
\bea
{\tilde V}
\approx (1+\frac{m}{\sqrt{3}H_0})
\frac{1}{\sqrt{2 \Omega}}
e^{-i\int^{t}dt' \Omega},
\quad
{\tilde H}_+
\approx (1-\frac{m}{\sqrt{3}H_0})
\frac{1}{\sqrt{2 \Omega}}
e^{-i\int^{t}dt' \Omega},
\label{wkb1}
\eea
where $\Omega$ is given by
\bea
&&\Omega^2
={\omega}^2
-\frac{1}{2}
\Big(\frac{{\Omega}_{,tt}}{{\Omega}}
-\frac{3}{2}
\frac{{\Omega}_{,t}^2}{{\Omega}^2}
\Big),
\label{freq_eq}
\eea
and satisfies $\Omega\to k_1$ for $t\to-\infty$.
On the other hand,
\bea
{\tilde H}_{\times}
\approx
\frac{1}{\sqrt{2 \Omega}}
e^{-i\int^{t}dt' \Omega}.
\eea

By using $b\approx (-3H_0 t)^{-\frac{1}{3}}$
and
$-H_0 \eta\approx (-3H_0t)^{\frac{1}{3}}$,
since the mixing is eventually vanishes,
the late time solutions are given in terms of the de Sitter mode solutions
by
\bea
\label{hiko}
{\tilde Y}\approx Y
&=&
\frac{A_{\tilde Y}}{\sqrt{2k}}
e^{i\frac{k}{H_0}(-3H_0t)^{\frac{1}{3}}}
\Big[(-3H_0 t)^{\frac{1}{3}}+\frac{i H_0}{k}\Big]
+
\frac{B_{\tilde Y}}{\sqrt{2k}}
e^{-i\frac{k}{H_0}(-3H_0t)^{\frac{1}{3}}}
\Big[(-3H_0t)^{\frac{1}{3}}-\frac{i H_0}{k}\Big].
\eea

We now match the WKB solution to the de Sitter solutions.
The details of the matching are summarized in Appendix A.
Keeping the accurary to the second order in adiabatic
approximation,
we finally obtain the power spectrum including
the leading order corrections from the direction dependent part
\bea
P_{\times}
&=&P_{\times}^{(0)}
\Big[
 1
+Q(r_2) \Big(\frac{H_0}{k}\Big)^{\frac{3}{2}}
 \cos\Big(2\sqrt{\frac{k}{H_0}}\Big)
+O\Big(\Big(\frac{H_0}{k}\Big)^2\Big)
\Big],
\nonumber\\
P_{V}
&=&P_{V}^{(0)}
\Big[
 1
+\frac{2m}{\sqrt{3}H_0}
+Q(r_2) \Big(\frac{H_0}{k}\Big)^{\frac{3}{2}}
 \cos\Big(2\sqrt{\frac{k}{H_0}}\Big)
+O\Big(\Big(\frac{H_0}{k}\Big)^2\Big)
\Big],
\nonumber\\
P_{+}
&=&P_{+}^{(0)}
\Big[
 1
-\frac{2m}{\sqrt{3}H_0}
+Q(r_2) \Big(\frac{H_0}{k}\Big)^{\frac{3}{2}}
 \cos\Big(2\sqrt{\frac{k}{H_0}}\Big)
+O\Big(\Big(\frac{H_0}{k}\Big)^2\Big)
\Big],
\eea
where $Q(r_2):=\frac{2}{3}-r_2^2$ ($r_2:=\frac{k_2}{k}$)
denotes the leading order corrections due to the anisotropy.
Thus,
the leading order correction appears at the third adiabatic order and
mode mixing gives the contribution of order $\frac{m}{H_0}$,
which results in a chiral asymmetry between two tensor polarizations.


\section{Planar modes $H_0<k_1\ll k_2$}

For the planar modes of $k_1\ll k_2$,
the WKB approximation is violated in the early times.
Instead,
we make use of approximate solutions.
We expand the effective frequency as,
\bea
\label{omega}
&&\omega_{\times}^2\approx
2^{\frac{4}{3}}
\big(k_1^2+ \delta \omega_{\times}^2e^{6H_0t}\big),
\quad
\omega_{11}^2\approx
2^{\frac{4}{3}}
\big(k_1^2+ \delta \omega_{11}^2 e^{6H_0t}\big),
\nonumber\\
&&\omega_{22}^2\approx 2^{\frac{4}{3}}
\big(k_1^2+ \delta \omega_{22}^2e^{6H_0t}\big),
\quad
\omega_{12}^2\approx 2^{\frac{4}{3}} (\delta \omega_{12}^2)e^{6H_0t}.
\eea
where
\bea
\label{do}
\delta \omega_{\times}^2
&=&k_2^2+\frac{4}{3}k_1^2+\frac{9H_0^2 k_2^2}{2^{\frac{1}{3}}k_1^2},
\quad
\delta\omega_{11}^2
=k_2^2+\frac43k_1^2 -2^{2/3} m^2\frac{4k_1^2-3k_2^2}{4k_1^2+3k_2^2}%
,\nonumber\\
\delta\omega_{22}^2
&=& k_2^2+\frac43k_1^2+\frac{18\times 2^{2/3} H_0^2 k_2^2}{4k_1^2+3k_2^2},
\quad\delta\omega_{12}^2
=-\frac{6 \times 2^{2/3} \sqrt{3}H_0 m k_2^2}
        {4k_1^2+3k_2^2}.
\eea

For convenience, we introduce
the {\it planarity} parameter $s$ ($\ll 1$),
defined by $k_2=\frac{k_1}{s}$.
Then, $\delta\omega$s can be expanded around $s=0$
as
\bea
\delta \omega_{\times}^2
&=&\frac{k_1^2+\frac{9H_0^2}{2^{\frac{1}{3}}}}{s^2}
+\frac{4}{3}k_1^2+O(s^2),
\quad
\delta\omega_{11}^2
=\frac{k_1^2}{s^2}
+\Big(\frac{4}{3}k_1^2+2^{\frac{2}{3}}m^2\Big)
+O(s^2),
\nonumber\\
\delta\omega_{22}^2
&=&\frac{k_1^2}{s^2}
+\Big(\frac{4}{3}k_1^2+6 \times 2^{\frac{2}{3}}H_0^2\Big)
+O(s^2),
\quad
\delta\omega_{12}^2
=-2^{\frac{5}{3}}\sqrt{3} H_0 m+O(s^2) ,
\eea
and
\bea
\delta {\tilde \omega}_{11}^2
&= &\frac{k_1^2}{s^2}
+\frac{4}{3}k_1^2
+2^{\frac{2}{3}}\times 3
 H_0^2
 \Big(
  1-\sqrt{1+\frac{m^2}{H_0^2}+\frac{m^4}{36H_0^4}}
 \Big)
+2^{-\frac{1}{3}}m^2
+O(s^2),
\nonumber\\
\delta {\tilde \omega}_{22}^2
&= &\frac{k_1^2}{s^2}
+\frac{4}{3}k_1^2
+2^{\frac{2}{3}}\times 3
 H_0^2
 \Big(
  1+\sqrt{1+\frac{m^2}{H_0^2}+\frac{m^4}{36H_0^4}}
 \Big)
+2^{-\frac{1}{3}}m^2
+O(s^2).
\eea
In the limit of $s\to 0$
the frequency diverges and
we focus on $s\neq 0 $.
Note that
$$
 \delta{\tilde\omega}_{11}^2
-\delta{\tilde \omega}_{22}^2
= -6 \times 2^{\frac{2}{3}}H_0^2
\sqrt{1+\frac{m^2}{H_0^2}+\frac{m^4}{36H_0^4}}
\simeq -6 \times 2^{\frac{2}{3}}H_0^2<0.
$$
In the small $s$ limit,
$\omega_{\tilde Y}^2$ behave as
\bea
\omega_{\tilde Y}^2&=&2^{\frac{2}{3}}k_2^2
          \sinh^{\frac{4}{3}}\Big(\frac{3H_0\tau}{2}\Big)
\sinh^{\frac{2}{3}}\big(3H_0\tau\big)
+O(s^0)
=\frac{2^{\frac{4}{3}}k_2^2 e^{6H_0 t}}
        {(1-e^{6H_0 t})^{\frac{4}{3}}}
+O(s^0),
\label{bob}
\eea
where the species of perturbations are irrelevant.

\subsection{$H_{\times}$ mode and power spectrum}

For $H_0t\ll -1$, the solutions for $H_\times$ are given by
\bea \label{inc2}
&&H^{(1)}_{\times}=\sqrt{\frac{\pi}{6H_0 \sinh (\pi q_1)}}
   J_{-iq_1} \Big(q_{\times}e^{3H_0t} \Big),
\eea
where
$q_{1}:=
\frac{2^{\frac{2}{3}}k_1}
     {3H_0}$
and
$q_{\times}:=
\frac{2^{\frac{2}{3}}(\delta \omega_{\times}^2)^{\frac{1}{2}}}
     {3H_0}$.
Note that
this solution reproduces the correct normalization
in the limit $t\to -\infty$.

When the WKB solution becomes valid,
we can match
the solution Eq. (\ref{inc2})
to the WKB solution.
Then, when the universe enters into the de Sitter phase,
we can match the WKB solution
to the de Sitter mode functions.
The details
are summarized in Appendix B 1.
In this subsection,
we show the final power spectrum of $H_{\times}$
for the planar modes
\bea
P_{h_{\times}}
&\simeq&
P_{h_{\times}}^{(0)}
\Big(
\coth\big(\pi q_1\big)
-\frac{\sin(2\Psi_{\times})}{\sinh(\pi q_1)}
\Big),
\eea
where
$$
\Psi_{\times} =
\frac{\sqrt{\pi}\Gamma(\frac{1}{3})}{3\times 2^{\frac{1}{3}}\Gamma(\frac{5}{6})}\frac{k}{H_0}
+O(\sqrt{\frac{k}{H_0}}).
$$
Here we use the fact that $k_2 \simeq k$ in the case of planar modes.

\begin{figure}[h]
\begin{center}
\includegraphics[width=.5\linewidth]{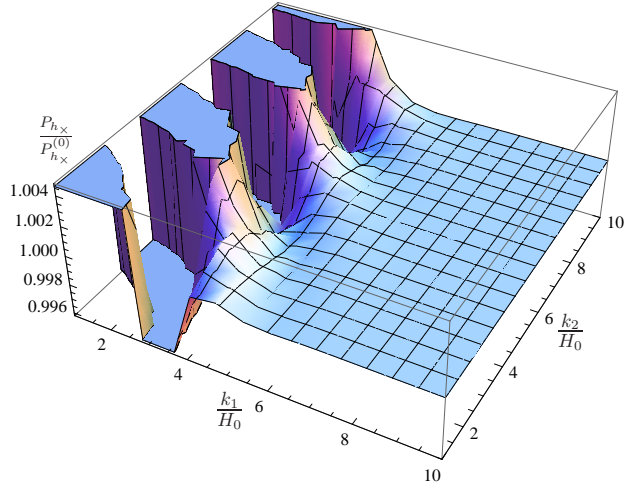}
\end{center}
\caption{Power spectrum of the tensor mode $H_\times$ relative to the
isotropic case.
The high momentum mode corresponds to large $k_1/H_0$ and $k_2/H_0$. 
The planar mode corresponds to the case that $k_1/ H_0$ 
becomes $O(1)$.} \label{fig:spec}
\end{figure}

As shown in Fig.~\ref{fig:spec}, the power spectrum exponentially approach to that in the isotropic case. 
On the other hand, for the planar mode with $k_1/H_0$ be $O(1)$, it leaves observable effect in the CMB. 


\subsection{Mixed modes and power spectra}

The evolution of $V$ and $H_{+}$ modes can be dealt with the parallel way.
The evolution equations at the initial times
are given in terms of the mixed form
\begin{eqnarray}
\frac{d^2}{dt^2}
\left(\begin{tabular}{c}
                     $V$ \\
                     $H_+$ \\
                   \end{tabular}\right)
+ 2^{4/3}  \left[ k_1^2{\bf 1} + \left(\begin{tabular}{cc}
                     $\delta\omega_{11}^2$& $ \delta\omega_{12}^2$ \\
                     $\delta \omega_{12}^2$ & $\delta \omega_{22}^2$\\
                   \end{tabular}\right)e^{6Ht}\right]
\left(\begin{tabular}{c}
                     $V$ \\
                     $H_+$ \\
                   \end{tabular}\right) =0,  \label{ddM2b}
\end{eqnarray}
where $\delta\omega^2$s are defined in Eq. (\ref{do}).
We may diagonalize the equation~(\ref{ddM2b}) by using the transformation
$$
\left(\begin{tabular}{c}
                     $V$ \\
                     $H_+$ \\
                   \end{tabular}\right) = O \left(\begin{tabular}{c}
                     $v$ \\
                     $h_+$ \\
                   \end{tabular}\right) = \left(\begin{tabular}{cc}
                     $\cos \psi$& $ -\sin \psi$ \\
                     $\sin \psi$ & $\cos \psi$ \\
                   \end{tabular}\right) \left(\begin{tabular}{c}
                     $v$ \\
                     $h_+$ \\
                   \end{tabular}\right)  .
$$
By choosing $\psi$ to be
\bea
\cos\psi
&=&\Big[\frac{1}{2}\Big(1+\frac{1}{\sqrt{1+\chi_0^2}}\Big)\Big]^{\frac{1}{2}},
\quad
\sin\psi
=
\Big[\frac{1}{2}\Big(1-\frac{1}{\sqrt{1+\chi_0^2}}\Big)\Big]^{\frac{1}{2}},
\nonumber\\
\chi_0
&:=&\frac{2\delta \omega_{12}^2}
{\delta \omega_{22}^2-\delta \omega_{11}^2}.
\label{psi}
\eea
At the initial times, $\chi_0$ becomes
$$
\chi_0 = -\frac{2 m/H_0}{\sqrt{3}(1-m^2/(6H_0^2))} +O(1/s^2).
$$

Then the frequency term can be diagonalized
$$
O^{-1}\left(\begin{tabular}{cc}
                     $\delta\omega_{11}^2$& $ \delta\omega_{12}^2$ \\
                     $\delta \omega_{12}^2$ & $\delta \omega_{22}^2$
                     \\
                   \end{tabular}\right)O
=   \left(\begin{tabular}{cc}
                 $\delta\tilde{\omega}{}_{11}^2$& $0$ \\
                 $0$ & $\delta\tilde{\omega}{}_{22}^2$\\
                   \end{tabular}\right),
$$
where
\begin{eqnarray}
\delta{\tilde \omega}_{11}^2
&=&\frac{1}{2}
\Big( \delta \omega_{11}^2+\delta \omega_{22}^2
  -\big(  \delta \omega_{22}^2-\delta \omega_{11}^2\big)
   \sqrt{1+\frac{4(\delta\omega_{12}^2)^2} {(\delta \omega_{11}^2
 -\delta \omega_{22}^2)^2} }
\Big),
\nonumber\\
\delta{\tilde \omega}_{22}^2
&=&\frac{1}{2}
\Big(
 \delta \omega_{11}^2
+\delta \omega_{22}^2
+\big(
  \delta \omega_{22}^2
 -\delta \omega_{11}^2
  \big)
\sqrt{1+
  \frac{4(\delta\omega_{12}^2)^2}
        {(\delta \omega_{11}^2
 -\delta \omega_{22}^2)^2}
   }
\Big).
\end{eqnarray}
The original mixed equations are now diagonalized for the new variables
\bea
\M=\left(\begin{tabular}{c}
                     $v$ \\
                     $h_+$ \\
                   \end{tabular}\right)
=\left(\begin{tabular}{c}
                     $V\cos\psi+ H_+\sin\psi$ \\
                     $H_+\cos\psi-V \sin\psi $ \\
                   \end{tabular}\right)\,.
\eea
The solutions are given by
\begin{eqnarray}
v
&\approx& \Big(1
+\frac{m}{\sqrt{3}H_0}\Big)
\sqrt{\frac{\pi}{6H_0 \sinh (\pi q_1)}}
   J_{-iq_1}
\Big(
q_{11}e^{3H_0t}
\Big),
\nonumber\\
h_{+}
&\approx&
\Big(1
-\frac{m}{\sqrt{3}H_0}\Big)\sqrt{\frac{\pi}{6H_0 \sinh (\pi q_1)}}
   J_{-iq_1}
\Big(
q_{22}e^{3H_0 t}
\Big),
\end{eqnarray}
where
$q_1:=\frac{2^{\frac{2}{3}}|k_1|}{3H_0}$,
$q_{11}:=\frac{2^{\frac{2}{3}}(\delta {\tilde \omega}_{11}^2)^{\frac{1}{2}}}
{3H_0}$,
and
$q_{22}:=\frac{2^{\frac{2}{3}}(\delta {\tilde \omega}_{22}^2)^{\frac{1}{2}}}
{3H_0}$,
and similarly to the non-planar high momentum modes,
factors of order $\frac{m}{H_0}$
represent corrections due to the initial mixing of modes
and the early time limit of the above solutions
correctly reproduces the normalized amplitude.
Note that
these solutions reproduce the correct normalizations
in the limit $t\to -\infty$.
By definition,
$V$ and $H_+$ are now given by
\bea
V&=&
\sqrt{\frac{\pi}{6H_0 \sinh (\pi q_1)}}
\Big[
\Big(1
+\frac{m}{\sqrt{3}H_0}\Big)
\cos\psi
J_{-iq_1}
\Big(
q_{11}e^{3H_0t}
\Big)
\nonumber\\
&-&
\Big(1
-\frac{m}{\sqrt{3}H_0}\Big)
\sin \psi
J_{-iq_1}
\Big(
q_{22}e^{3H_0t}
\Big)
\Big],
\nonumber\\
H_{+}
&=&\sqrt{\frac{\pi}{6H_0 \sinh (\pi q_1)}}
\Big[
\Big(1
+\frac{m}{\sqrt{3}H_0}\Big)
\sin\psi
J_{-iq_1}
\Big(
q_{11}e^{3H_0t}
\Big)
\nonumber\\
&+&
\Big(1
-\frac{m}{\sqrt{3}H_0}\Big)
\cos \psi
J_{-iq_1}
\Big(
q_{22}e^{3H_0t}
\Big)
\Big].\label{bf_wkb}
\eea

When the WKB becomes valid,
the coupling between $\tilde V$ and ${\tilde H}_+$
is negligible.
At this time,
we can match the solution Eq. (\ref{bf_wkb}) to WKB solutions.
Then,
when the universe enters into the de Sitter phase,
we finally match the WKB solutions,
to the de Sitter mode functions.
The essential procedure is almost the same
as the case of the $H_{\times}$ mode.
The details are summarized in the Appendix B 2.
In this subsection we only show the final results:
The power spectrum of $\tilde V$ ($=V$) for the planar modes
is
given by
\bea
P_{V}
&\simeq&
P_{V}^{(0)}
\Big(1
+\frac{2m}{\sqrt{3}H_0}\Big)
\Big(
\coth(\pi q_1)
-\frac{\sin(2\Psi_{V,1})}{\sinh(\pi q_1)}
\Big),
\quad
\eea
where 
$$
\Psi_{V,1}
=
\frac{\sqrt{\pi}\Gamma(\frac{1}{3})}{3\times 2^{\frac{1}{3}}\Gamma(\frac{5}{6})}\frac{k}{H_0}
+O(\sqrt{\frac{k}{H_0}}).
$$
Similarly,
the power spectrum of ${\tilde H}_{+}$
($=H_{+}$) for the planar modes is
given by
\bea
P_{h_+}
&\simeq&
P_{h_+}^{(0)}
\Big(1
-\frac{2m}{\sqrt{3}H_0}\Big)
\Big(
\coth(\pi q_1)
-\frac{\sin(2\Psi_{+,2})}{\sinh(\pi q_1)}
\Big),
\nonumber\\
\Psi_{+,2}
&:=&
\frac{\sqrt{\pi}\Gamma(\frac{1}{3})}{3\times 2^{\frac{1}{3}}\Gamma(\frac{5}{6})}\frac{k}{H_0}
+O(\sqrt{\frac{k}{H_0}}).
\eea
Note that the forms of the power spectra for the $H_+$ and $V$ modes are almost the same as that of the $H_\times$ mode except for the global scaling due to the initial mode mixing.


\section{Conclusion}

In this paper,
we have analytically investigated the corrections
to the power spectra of the cosmological perturbations
due to the preinflationary anisotropy of the universe,
for the high momentum modes
in the context of WKB approximations.
The first motivation to consider the anisotropic universe
is that
even if the present universe is almost isotropic,
it does not mean that it is also isotropic from the beginning.
It would be more generic that the initial universe is highly anisotropic.
The second motivation comes from observations.
In recent years,
several groups have reported the so-called low-$\ell$ anomalies
in large angular power of CMB fluctuations.
They may
be produced by the breaking of the rotational symmetry
in the early universe.

We considered the Einstein gravity coupled to a
light scalar field.
We assumed that this scalar field initially stays at a
very large (super-Planck) field value,
and then starts to roll down slowly.
If the mass of the scalar field is small enough,
the kinetic energy of the scalar field does not
affect the spacetime dynamics significantly.
Imposing the regularity of the spacetime at the initial time,
one of two planar branches of the Kasner-de Sitter solution,
whose initial geometry becomes a (Milne) patch of
the Minkowski spacetime,
is a good approximation for
describing the cosmic isotropization.
The cosmic isotropization
takes place within a few Hubble times.
During the subsequent inflationary stage,
the scalar field rolls down toward the true minimum
and plays the role of the inflaton.

Then, we investigated the analytic expressions
for the spectra of the cosmological perturbations
produced in the above anisotropic background
for the high momentum modes.
In the anisotropic background,
there are 2 scalar and 1 vector modes
in terms of the two-dimensional flat space.
In the isotropic limit,
this vector mode in the anisotropic universe
reduces to one of tensor polarizations
in the flat 3-dimensional space,
while two scalar modes reduce to
one scalar and the other tensor polarization there.
During the anisotropic phase,
two scalar modes are coupled,
resulting in the asymmetry between
the spectra of two tensor polarizations
obtained in the isotropic limit.
Since at the initial times, the coupling is absent,
we could define the adiabatic vacuum,
which we call the anisotropic vacuum,
and canonically quantize the perturbations.
Our anisotropic vacuum is definitely different from
the standard Bunch-Davis vacuum.
In addition, the presence of the anisotropic vacuum
is specific to
our particular choice of
the Kasner parameter.

For the non-planar modes $k,k_1\gg H_0$,
for the sufficiently high momentum
the WKB approximation is valid but
the mixing angle does not vanish
at the early time.
At the leading order,
the power spectra of all the perturbation modes
contain
the corrections due to the
nonstandard propagation in the
anisotropic background,
and the effects of the initial mixing of modes.
The former has the universal form,
while the latter induces an asymmetry
of two gravitational wave polarizations in the
isotropic limit.
The modifications of spectra
appear in the oscillatory behaviors
of the primordial spectrum.
On the other hand,
for the planar mode, i.e., $k\gg k_1 \sim H_0$,
although the WKB approximation is broken at the very early times,
the mode mixing does not take place significantly.
For the modes of $k_1=0$, the adiabaticity parameter initially
diverges and hence the anisotropic vacuum is not well defined.
However, such modes are not relevant for the observations.

One of the main results which we have obtained is
that, irrespective of the non-planar or planar modes,
for the high momentum modes
the ratio of the power spectra between
two tensor polarizations is given by
\bea
\frac{P_{h_+}}{P_{h_\times}}\approx  1
-\frac{2m}{\sqrt{3}H_0}.
\eea
In the chaotic inflation typically $\frac{m}{H_0}=O(0.1)$,
and hence there is a difference of
the spectra of two gravitational wave polarizations,
which is of roughly ten percent.

Before closing this article,
it may be important to mention
the effects of the primordial anisotropy
on the higher order spectra,
in particular on the bispectrum,
and
the primordial non-Gaussianities.
In the model discussed in this paper
the initial vacuum is not
the standard Bunch-Davis vacuum,
and the subsequent evolution is almost the same
as that in the single field, slow-roll inflation.
Thus, we expect that
the folded shape bispectrum
where three momenta satisfy $k_1+ k_2 \sim k_3$
would become dominant
\cite{chen,chks} (see also \cite{isca}),
while the local shape bispectrum of
$k_2\sim k_3\gg k_1$
would be negligible \cite{jm}.
However,
the particular local type bispectrum 
where $k_2$ and $k_3$ almost lie in the plane
of the $y$ and $z$ directions of Eq. (\ref{metric})
while $k_1$ is nearly orthogonal to this plane ($k_2\sim k_3\gg k_1$),
could be much different from the case of the isotropic universe,
since 
as in the case of the spectra for the planar modes
it could be sensitive to the anisotropy.
For the non-planar case of $k_1+ k_2 \sim k_3$,
the amount of the non-Gaussianities would be determined
by the coefficient $B_{\tilde V}$ in Eq. (\ref{hiko})
which represents
the amplitude of the negative frequency mode;
\bea
{\rm Re} (B_{\tilde V})\sim
 Q(r_2) \Big(\frac{H_0}{k}\Big)^{\frac{3}{2}}
 \sin\big(\sqrt{\frac{k}{H_0}}+\phi_0\big),
\eea
where $\phi_0$ denotes some constant phase.
Note that due to the effect of the initial mode mixing
${\rm Re} (B_{\tilde V})$ may be amplified by some factor of order
$\frac{m}{H_0}\sim 0.1$.
Therefore,
the bispectrum could exhibit an oscillatory behavior,
and contain the information on the primordial anisotropy
through the factor $Q(r_2)$,
which may be distinguishable if detected.
The concrete evaluation of the bispectrum
in the anisotropic universe
and the detectability of the non-Gaussianities
will be interesting issues and
should be investigated in the future studies.

\begin{acknowledgments}
HCK was supported in part by the Korea Science and Engineering Foundation
(KOSEF) grant funded by the Korea government (MEST) (No.2010-0011308).
We are grateful for discussions
during the COSMO/COSPA 2010 held in Japan.
MM is also grateful for the hospotality
of the Center for Quantum Spacetime (Sogang University)
and the Chungju National University.
\end{acknowledgments}
 \vspace{1cm}

\appendix

\section{On the power spectra for the non-planar, high-momentum modes}

We will summarize the matching of the WKB and de Sitter mode functions
for the non-planar high momentum modes,
given by Eqs. (\ref{wkb1}) and (\ref{hiko}), respectively.

Keeping the accuracy to the second order in adiabatic approximation,
the solution to Eq. (\ref{freq_eq}) is given by
\bea
\Omega=\omega(1+\delta),
\eea
where
$$
\delta:=-\frac{1}{4\omega^2}
\Big(\frac{\omega_{,tt}}{\omega}
-\frac{3\omega_{,t}^2}{2\omega^2}\Big).
$$
Defining
\bea
a:=1-\frac{\delta}{2},\quad
b:=1+\frac{\delta}{2},\quad
c:=-\frac{\omega_{,t}}{2\omega^2}
 \Big(1-\frac{\delta}{2}+\frac{\omega\delta_{,t}}{\omega_{,t}}\Big),
\eea
and using the results of
matching at some appropriate time $t=t_{\ast}$,
the final power spectrum is given by
\bea
P_{\tilde Y}&=&P_{\tilde Y}^{(0)}
\frac{K_{Y}}{\nu_{\ast}}
\Big\{
b^2\nu_{\ast}^2
\Big(
\cos\big[\frac{k}{H_0}(-3H_0t_{\ast})^{\frac{1}{3}}\big]
-\frac{H_0}{k}\frac{1}{(-3H_0t)^{\frac{1}{3}}}
\sin\big[\frac{k}{H_0}(-3H_0t_{\ast})^{\frac{1}{3}}\big]
\Big)^2
\nonumber\\
&+&
\Big(
-\big(a
+c \frac{H}{k}\frac{\nu_{\ast}}{(-3H_0t_{\ast})^{\frac{1}{3}}}
\big)
\sin\big[\frac{k}{H_0}(-3H_0t_{\ast})^{\frac{1}{3}}\big]
+c\nu_{\ast}
\cos \big[\frac{k}{H_0}(-3H_0t_{\ast})^{\frac{1}{3}}\big]
\Big)^2
\Big\},
\eea
where
$\nu_{\ast}:=\frac{\omega_{\ast}}{k}(-3H_0t_{\ast})^{\frac{2}{3}}$
and $\Phi=\int^{t_{\ast}} dt' \omega(t')$,
and $P_{Y}^{(0)}$ is given by \eqref{zeroth}.
$K_{\tilde Y}$ represents the contribution of
the initial mixing of modes,
which are given by
$K_\times =1$, $K_V=1+\frac{2m}{\sqrt{3}H_0}$
and $K_+=1-\frac{2m}{\sqrt{3}H_0}$.

For all ${\tilde  H}_{+}$, ${\tilde V}$ and $H_{\times}$ modes,
we obtain in the high momentum limit,
\bea
\omega_{\ast}^2= k^2 e^{4\alpha}
\Big(1+2e^{-3\alpha}Q(r_2)\Big)
+O(m^2),
\eea
where $Q(r_2):=\frac{2}{3}-r_2^2$ ($r_2:=\frac{k_2}{k}$)
denotes the leading order corrections due to the anisotropy.
Note that $-\frac{1}{3}\leq Q(r_2) \leq \frac{2}{3}$
and $O(m^2)$ terms contain the slow-roll corrections.
Thus, as long as our concerns are in the high momentum modes,
$O(m^2)$ terms are not important.
The WKB approximations are valid as long as
the adiabaticity parameter Eq (\ref{ups})
is smaller than unity, 
namely $-H_0t \gg \Big(\frac{H_0}{k}\Big)^3$.
On the other hand, the late time approximation is valid if $e^{\alpha}\gg 1$.
The matching time $t=t_{\ast}$ should be chosen so that the error is minimized and
is given in terms of their geometric mean~\cite{km}:
\bea
\label{mika2}
e^{\alpha}\Big|_{t=t_{\ast}}
:= \Big(\frac{k}{H_0}\Big)^{\frac{1}{2}}.
\eea
At $t=t_*$,
$\epsilon_{{\tilde Y},\ast}\simeq \frac{H_0^{\frac{1}{2}}}{k^{\frac{1}{2}}}$.
Making use of
$
(-3H_0t_{\ast})^{\frac{1}{3}}
=\Big(\frac{H_0}{k}\Big)^{\frac{1}{2}}
\Big[
1+O(\frac{k^3}{H_0^3})
\Big].
$
the power spectrum is given by
\bea
P_{\tilde Y}&=&
P_{\tilde Y}^{(0)}
\frac{K_Y}{\nu_{\ast}}
\Big\{
b^2\nu_{\ast}^2
\Big(
\cos\sqrt{\frac{k}{H_0}}
-\sqrt{\frac{H_0}{k}}
\sin\sqrt{\frac{k}{H_0}}
\Big)^2
\nonumber\\
&+&
\Big(
-\big(
a
+c \nu_{\ast}\sqrt{\frac{H_0}{k}}\big)
\sin\sqrt{\frac{k}{H_0}}
+c\nu_{\ast}
\cos \sqrt{\frac{k}{H_0}}
\Big)^2
\Big\}.
\eea
As shown in the previous section, $\sin \theta$ is $O(m/H_0)$ at early times and vanishes at late times, and thus
the initial condition for $V$ and $H_+$ modes will be specified by using the WKB approximation for $\tilde V$ and $\tilde H_+$.
Since $\tilde V=V$ and $\tilde H_+ =H_+$ at $t\to 0-$,
the power spectra for $\tilde Y$
reproduce those of $Y$.
Therefore, $P_{Y}=P_{\tilde Y}$.

Then,
evaluating at $t=t_{1}$
up to $O\Big(\big(\frac{H_0}{k}\big)^2\Big)$,
\bea
&&
\nu_{\ast}
=1+Q(r_2)\Big(\frac{H_0}{k}\Big)^{\frac{3}{2}}
+O\Big(\Big(\frac{H_0}{k}\Big)^2\Big),
\quad
a=1+\frac{H_0}{2k}+O\Big(\Big(\frac{H_0}{k}\Big)^2\Big),
\nonumber\\
&&b=1-\frac{H_0}{2k}+O\Big(\Big(\frac{H_0}{k}\Big)^2\Big),
\quad
c=-\sqrt{\frac{H_0}{k}}\Big(1-\frac{H_0}{2k}\Big)
+O\Big(\Big(\frac{H_0}{k}\Big)^2\Big).
\eea
Therefore,
we obtain the power spectrum including
the leading order corrections from the direction dependent part
\bea
P_{\times}
&=&P_{\times}^{(0)}
\Big[
 1
+Q(r_2) \Big(\frac{H_0}{k}\Big)^{\frac{3}{2}}
 \cos\Big(2\sqrt{\frac{k}{H_0}}\Big)
+O\Big(\Big(\frac{H_0}{k}\Big)^2\Big)
\Big],
\nonumber\\
P_{V}
&=&P_{V}^{(0)}
\Big[
 1
+\frac{2m}{\sqrt{3}H_0}
+Q(r_2) \Big(\frac{H_0}{k}\Big)^{\frac{3}{2}}
 \cos\Big(2\sqrt{\frac{k}{H_0}}\Big)
+O\Big(\Big(\frac{H_0}{k}\Big)^2\Big)
\Big],
\nonumber\\
P_{+}
&=&P_{+}^{(0)}
\Big[
 1
-\frac{2m}{\sqrt{3}H_0}
+Q(r_2) \Big(\frac{H_0}{k}\Big)^{\frac{3}{2}}
 \cos\Big(2\sqrt{\frac{k}{H_0}}\Big)
+O\Big(\Big(\frac{H_0}{k}\Big)^2\Big)
\Big].
\eea

\section{On the power spectra of the planar modes}

\subsection{On $H_\times$}

Here,
we explain the derivation of the power spectrum of $H_\times$
for the planar modes.

For $H_0t\ll -1$, the solutions for $H_\times$ are given by
\bea \label{inc}
&&H^{(1)}_{\times}=\sqrt{\frac{\pi}{6H_0 \sinh (\pi q_1)}}
   J_{-iq_1} \Big(q_{\times}e^{3H_0t} \Big),
\eea
where
$q_{1}:=
\frac{2^{\frac{2}{3}}k_1}
     {3H_0}$
and
$q_{\times}:=
\frac{2^{\frac{2}{3}}(\delta \omega_{\times}^2)^{\frac{1}{2}}}
     {3H_0}$.
Note that
this solution reproduces the correct normalization
in the limit $t\to -\infty$.
During the period of time $2t_{1,\times}< t\ll 
-\frac{H_0^2}{k^3}$, where $t_1$ is the matching time given below,
the WKB approximation
\bea
H^{(2)}_{\times}
=\frac{B_{\times,+}}{\sqrt{2\omega_\times}}{\rm exp}
\Big[
-i\int_{t_{1,\times}}^t dt' \omega_{\times}(t')
\Big]
+\frac{B_{\times,-}}{\sqrt{2\omega_\times}}{\rm exp}
\Big[
i\int_{t_{1,\times}}^t dt' \omega_{\times}(t')
\Big],\label{WKB2}
\eea
holds well.
In order
for the error to be minimized, 
we need to 
choose the matching time of
the two solutions~(\ref{inc}) and (\ref{WKB2}),
$t=t_{1,\times}$ to be
\bea
e^{3H_0t_{1,\times}}
\simeq \frac{3}{2^{\frac{2}{3}}}
\sqrt{\frac{H_0}
{(\delta\omega_{\times}^2)^{\frac{1}{2}}}},
\eea
since the adiabaticity parameter behaves as
$
\epsilon_{\times}\simeq
\frac{3\times 2^{\frac{1}{3}}H_0 }
     {(\delta\omega_{\times}^2)^{\frac{1}{2}}}
e^{-3H_0t}$.
Since 
$q_{\times}e^{3H_0t_{1,\times}}\gg 1$ at $t=t_1$, 
\bea
H^{(1)}_{\times}\simeq
\sqrt{\frac{1}{
3H_0 q_{\times}
e^{3H_0 t}
 \sinh (\pi q_1)}}
   \sin\Upsilon_{\times},
\label{yoko}
\eea
where
$\Upsilon_{\times}
= q_{\times}e^{3H_0 t}
+\frac{\pi}{4}+\frac{iq_1\pi}{2}$.
Noting
\bea
\frac{3H_0 q_\times e^{3Ht_{1,\times}}}
{\omega_{\times}|_{t=t_{1,\times}}}
=\Big(1+\frac{H_0 q_1^2}
     {\big(\delta\omega_{\times}^{2}\big)^{\frac{1}{2}}}\Big)^{-\frac{1}{2}},
\eea
we obtain
\bea
B_{\times,+}&=&\sqrt{\frac{\omega_\times|_{t=t_{1,\times}}}{2}}
   \Big( H_{\times}^{(1)}
 + i\frac{H^{(1)}_{\times,t}}{\omega_\times}
\Big)_{t=t_{1,\times}}
\nonumber\\
&\simeq&
\frac{i}{\sqrt{2\sinh(\pi q_1)}}
\Big[
\Big(1+\frac{H_0 q_1^2}
     {\big(\delta\omega_{\times}^{2}\big)^{\frac{1}{2}}}\Big)^{-\frac{1}{4}}
\cos\Upsilon_{\times}|_{t=t_{1,\times}}
-i
\Big(1+\frac{H_0 q_1^2}
     {\big(\delta\omega_{\times}^{2}\big)^{\frac{1}{2}}}\Big)^{\frac{1}{4}}
 \sin\Upsilon_{\times}\big|_{t=t_{1,\times}}
\Big],
\nonumber\\
B_{\times,-}&=&\sqrt{\frac{\omega_\times|_{t=t_{1,\times}}}{2}}
   \Big( H_{\times}^{(1)}
 - i\frac{H^{(1)}_{\times,t}}{\omega_\times}\Big)_{t=t_{1,\times}}
\nonumber\\
&\simeq&
\frac{(-i)}{\sqrt{2\sinh(\pi q_1)}}
\Big[
\Big(1+\frac{H_0 q_1^2}
     {\big(\delta\omega_{\times}^{2}\big)^{\frac{1}{2}}}\Big)^{-\frac{1}{4}}
\cos\Upsilon_{\times}|_{t=t_{1,\times}}
+i
\Big(1+\frac{H_0 q_1^2}
     {\big(\delta\omega_{\times}^{2}\big)^{\frac{1}{2}}}\Big)^{\frac{1}{4}}
 \sin\Upsilon_{\times}\big|_{t=t_{1,\times}}
\Big].
\eea
Since
$q_1^2\ll\frac{H_0}{(\delta\omega_{\times}^{2})^{\frac{1}{2}}}$,
at the leading order 
they reduce to
\bea
B_{\times,+}
\simeq \frac{1}{(1-e^{-2\pi q_1})^{\frac{1}{2}}}
e^{i\big(\frac{\pi}{4}
-\sqrt{\frac{(\delta\omega_{\times}^2)^{\frac{1}{2}}}{H_0}}\big)},
\quad
B_{\times,-}
\simeq \frac{e^{-\pi q_1}}{(1-e^{-2\pi q_1})^{\frac{1}{2}}}
e^{-i\big(\frac{\pi}{4}
-\sqrt{\frac{(\delta\omega_{\times}^2)^{\frac{1}{2}}}{H_0}}\big)}.
\eea

Then, at $t=t_{\ast}$ ($|t_{\ast}|\ll |t_{1,\times}|$),
where $t_{\ast}$ is chosen to be Eq. (\ref{mika2})
as for the non-planar modes,
the WKB solution are matched to
the de Sitter mode function
\bea
H^{(3)}_\times
&=&
\frac{C_{\times,+}}{\sqrt{2k}}e^{i\frac{k}{H_0}(-3H_0t)^{\frac{1}{3}}}
\Big[\big(-3H_0 t\big)^{\frac{1}{3}}+\frac{iH_0}{k}\Big]
+
\frac{C_{\times,-}}{\sqrt{2k}}e^{-i\frac{k}{H_0}(-3H_0t)^{\frac{1}{3}}}
\Big[\big(-3H_0 t\big)^{\frac{1}{3}}-\frac{iH_0}{k}\Big].
\nonumber\\
&\xrightarrow[t\to 0-]{}&
\frac{iH_0}{\sqrt{2k^3}}
\Big(
C_{\times,+}-C_{\times,-}
\Big).
\eea
For the high momentum modes,
\bea
C_{\times,+}\simeq e^{-i\sqrt{\frac{k}{H_0}}}e^{-i\Phi}B_{\times,+},
\quad
C_{\times,-}\simeq e^{i\sqrt{\frac{k}{H_0}}}e^{i\Phi}B_{\times,-},
\eea
where $\Phi=\int_{t_{1,\times}}^{t_{\ast}}
dt' \omega_{\times}(t')$.
Thus,
\bea
C_{\times,+}-C_{\times,-}
=\frac{1}{(1-e^{-2\pi q_1})^{\frac{1}{2}}}
\Big\{
\big(1-e^{-\pi q_1}\big)
\cos\big(\frac{\pi}{4}-\Psi_{\times}\big)
+i\big(1+e^{-\pi q_1}\big)
\sin\big(\frac{\pi}{4}-\Psi_{\times}\big)
\Big\},
\eea
where
$\Psi_{\times}:=
\sqrt{
\frac{(\delta\omega^2_{\times})^{\frac{1}{2}}}
     {H_0}
}
+\Phi+\sqrt{\frac{k}{H_0}}$.
Therefore, the power spectrum for the $H_{\times}$ mode is
given by
\bea
P_{h_{\times}}
&\simeq&
P_{h_{\times}}^{(0)}
\Big(
\coth\big(\pi q_1\big)
-\frac{\sin(2\Psi_{\times})}{\sinh(\pi q_1)}
\Big).
\eea
Finally, we estimate $\Psi_{\times}$.
From Eq. \eqref{bob},
\bea
\Phi= \int_{t_{1,\times}}^{t_{\ast}}dt
\frac{2^{\frac{2}{3}}k_2 e^{3H_0t}}
   {(1-e^{6H_0t})^{\frac{2}{3}}}
=\frac{2^{\frac{2}{3}}k_2}{3H_0}
e^{3H_0t }{}_2 F_1\Big[\frac{1}{2},\frac{2}{3},\frac{3}{2}, e^{6H_0 t}\Big]
\Big|_{t_{1,\times}}^{t_{\ast}}
\simeq
\frac{\sqrt{\pi}\Gamma(\frac{1}{3})}{3\times 2^{\frac{1}{3}}\Gamma(\frac{5}{6})}\frac{k_2}{H_0},
\eea
and the $\Phi$ term dominates $\Psi_{\times}$:
$$
\Psi_{\times} =
\frac{\sqrt{\pi}\Gamma(\frac{1}{3})}{3\times 2^{\frac{1}{3}}\Gamma(\frac{5}{6})}\frac{
k}{H_0}
+O(\sqrt{\frac{
k}{H_0}}).
$$
Here we use the fact that $k_2 \simeq k$ in the case of planar modes.

\subsection{On $V$ and $H_{+}$}

Similarly, we explain the derivation of the power spectra
of $V$ and $H_+$ for the planar modes.
Firstly, we match the solutions Eq. (\ref{bf_wkb}) to the
WKB solutions,
and then match them to the mode functions in
the de Sitter inflation.
Most computations are the same as in the case of $H_{\times}$.

\subsubsection{$\tilde V$ mode}

At the time when the WKB becomes valid,
the coupling between $\tilde V$ and ${\tilde H}_+$
is negligible.
At $t=t_{1,V}$ given by
\bea
e^{3H_0t_{1,V}}
\simeq \frac{3}{2^{\frac{2}{3}}}
\sqrt{\frac{H_0}
{(\delta {\tilde \omega}_{11}^2)^{\frac{1}{2}}}},
\eea
the asymptotic form of the first solution
\bea
{\tilde V}&\simeq &
\sqrt{\frac{1}{e^{3H_0 t} \sinh (\pi q_1)}}
\Big[
\Big(1
+\frac{m}{\sqrt{3}H_0}\Big)
\frac{\cos\big(\psi-\theta\big)}
     {2^{\frac{1}{3}}(\delta {\tilde \omega}_{11}^2)^{\frac{1}{4}} }
   \sin \Upsilon_{11}
\nonumber\\
&-&\Big(1
-\frac{m}{\sqrt{3}H_0}\Big)
\frac{\sin\big( \psi -\theta \big)}
      {2^{\frac{1}{3}}(\delta {\tilde \omega}_{22}^2)^{\frac{1}{4}} }
   \cos \Upsilon_{22}
\Big],
\eea
where
$\Upsilon_{11}
=q_{11}e^{3H_0t}
+\frac{\pi}{4}+\frac{iq_1\pi}{2}$
and
$\Upsilon_{22}
=q_{22}e^{3H_0t}
+\frac{\pi}{4}+\frac{iq_1\pi}{2}$,
is matched to the WKB solution given by
\bea
{\tilde V}^{(2)}
=\frac{B_{V,+}}{\sqrt{2{\tilde \omega}_{11}}}{\rm exp}
\Big[
-i\int_{t_{1,V}}^t dt' {\tilde \omega}_{11}(t')
\Big]
+\frac{B_{V,-}}{\sqrt{2{\tilde \omega}_{11}}}{\rm exp}
\Big[
i\int_{t_{1,V}}^t dt' {\tilde \omega}_{11}(t')
\Big].
\eea
Since
from Eqs. (\ref{theta}) and (\ref{psi}), at $t=t_{1,V}$
\bea
\chi\Big|_{t=t_{1,V}}
=\frac{2\omega_{12}^2}{\omega_{22}^2-\omega_{11}^2}\Big|_{t=t_{1,V}}
\simeq \frac{2\delta \omega_{12}^2}{\delta\omega_{22}^2-\delta\omega_{11}^2}
=\chi_0,
\label{angle}
\eea
at the leading order,
we obtain $\psi\approx \theta_{1,V}$.
The first solution then reduces to
\bea
{\tilde V}&\simeq &
\sqrt{\frac{1}{e^{3H_0 t} \sinh (\pi q_1)}}
\Big(1
+\frac{m}{\sqrt{3}H_0}\Big)
\frac{1
     }
     {2^{\frac{1}{3}}(\delta {\tilde \omega}_{11}^2)^{\frac{1}{4}} }
   \sin \Upsilon_{11}.
\eea
Noting that
$q_{11}^2\ll\frac{(\delta{\tilde \omega}_{11}^2)^{\frac{1}{2}}}{H_0}$,
after matching these solutions,
we obtain
\bea
B_{V,+}
&\simeq &\frac{1}{(1-e^{-2\pi q_1})^{\frac{1}{2}}}
\Big(1
+\frac{m}{\sqrt{3}H_0}\Big)
e^{i\big(\frac{\pi}{4}
-\sqrt{\frac{(\delta{\tilde\omega}_{11}^2)^{\frac{1}{2}}}{H_0}}\big)},
\nonumber\\
B_{V,-}
&\simeq
&\frac{e^{-\pi q_1}}{(1-e^{-2\pi q_1})^{\frac{1}{2}}}
\Big(1
+\frac{m}{\sqrt{3}H_0}\Big)
e^{-i\big(\frac{\pi}{4}
-\sqrt{\frac{(\delta{\tilde\omega}_{11}^2)^{\frac{1}{2}}}{H_0}}\big)}.
\eea

Then, at $t=t_{\ast}$ ($|t_{\ast}|\ll |t_{1,V}|$),
where $t_{\ast}$ is chosen to be Eq. (\ref{mika2})
as for the non-planar high momentum modes,
the WKB solution are matched to
the de Sitter mode function
\bea
{\tilde V}^{(3)}&=&
\frac{C_{V,+}}{\sqrt{2k}}e^{i\frac{k}{H_0}(-3H_0t)^{\frac{1}{3}}}
\Big[\big(-3H_0 t\big)^{\frac{1}{3}}+\frac{iH_0}{k}\Big]
+
\frac{C_{V,-}}{\sqrt{2k}}e^{-i\frac{k}{H_0}(-3H_0t)^{\frac{1}{3}}}
\Big[\big(-3H_0 t\big)^{\frac{1}{3}}-\frac{iH_0}{k}\Big]
\nonumber\\
&\xrightarrow[t\to 0-]{}&
\frac{iH_0}{\sqrt{2k^3}}
\Big(
C_{V,+}-C_{V,-}
\Big).
\eea
For the high momentum modes,
\bea
C_{V,+}-C_{V,-}
\simeq  \frac{1
+\frac{m}{\sqrt{3}H_0}}
{(1-e^{-2\pi q_1})^{\frac{1}{2}}}
\Big\{
\big(1-e^{-\pi q_1}\big)
 \cos\big(\frac{\pi}{4}-\Psi_{V,1}\big)
+i\big(1+e^{-\pi q_1}\big)
 \sin\big(\frac{\pi}{4}-\Psi_{V,1}\big)
\Big\},
\eea
where
$\Psi_{V,1}:=
\sqrt{
\frac{(\delta{\tilde \omega}^2_{11})^{\frac{1}{2}}}
     {H_0}}
+\Phi+\sqrt{\frac{k}{H_0}}$
and
$\Psi_{V,2}:=
\sqrt{
\frac{(\delta{\tilde \omega}^2_{22})}
     {H_0(\delta{\tilde \omega}^2_{11})^{\frac{1}{2}}}}
+\Phi+\sqrt{\frac{k}{H_0}}$.
Thus, the power spectrum for the $\tilde V$ ($=V$) mode is
given by
\bea
P_{V}
&\simeq&
P_{V}^{(0)}
\Big(1
+\frac{2m}{\sqrt{3}H_0}\Big)
\Big(
\coth(\pi q_1)
-\frac{\sin(2\Psi_{V,1})}{\sinh(\pi q_1)}
\Big),
\quad
\eea
where the $\Phi$ term dominates $\Psi_{V,1}$ and
$$
\Psi_{V,1} =
\frac{\sqrt{\pi}\Gamma(\frac{1}{3})}{3\times 2^{\frac{1}{3}}\Gamma(\frac{5}{6})}\frac{k}{H_0}
+O(\sqrt{\frac{k}{H_0}}).
$$
Here we use the fact that $k_2 \simeq k$ in the case of planar modes.

\subsubsection{${\tilde H}_{+}$ mode}

Similarly for the ${\tilde H}_{+}$ mode,
at $t=t_{1,+}$ 
\bea
e^{3H_0t_{1,+}}
\simeq \frac{3}{2^{\frac{2}{3}}}
\sqrt{\frac{H_0}
{(\delta {\tilde \omega}_{22}^2)^{\frac{1}{2}}}},
\eea
the asymptotic form of the first solution
\bea
{\tilde H}_{+}
&\simeq &
\sqrt{\frac{1}{e^{3H_0t} \sinh (\pi q_1)}}
\Big[
\Big(1
+\frac{m}{\sqrt{3}H_0}\Big)
\frac{\sin\big(\psi-\theta\big)}
     {2^{\frac{1}{3}}(\delta {\tilde \omega}_{11}^2)^{\frac{1}{4}} }
   \sin \Upsilon_{11}
\nonumber\\
&+&
\Big(1
-\frac{m}{\sqrt{3}H_0}\Big)
\frac{\cos\big( \psi -\theta \big)}
      {2^{\frac{1}{3}}(\delta {\tilde \omega}_{22}^2)^{\frac{1}{4}} }
   \cos \Upsilon_{22}
\Big],
\eea
is matched to the WKB solution is given by
\bea
{\tilde H}_+^{(2)}
=\frac{B_{+,+}}{\sqrt{2{\tilde \omega}_{22}}}{\rm exp}
\Big[
-i\int_{t_{1,+}}^t dt' {\tilde \omega}_{22}(t')
\Big]
+\frac{B_{+,-}}{\sqrt{2{\tilde \omega}_{22}}}{\rm exp}
\Big[
i\int_{t_{1,+}}^t dt' {\tilde \omega}_{22}(t')
\Big].
\eea
From Eqs. (\ref{theta}) and (\ref{psi}), at $t=t_{{1,+}}$
the similar relation to Eq. (\ref{angle}) is obtained,
which leads to $\psi\approx \theta_{1,+}$.
Thus, the first solution reduces to
\bea
{\tilde H}_{+}
&\simeq &
\sqrt{\frac{1}{e^{3H_0t} \sinh (\pi q_1)}}
\Big(1
-\frac{m}{\sqrt{3}H_0}\Big)
\frac{1 
      }
      {2^{\frac{1}{3}}(\delta {\tilde \omega}_{22}^2)^{\frac{1}{4}} }
   \cos \Upsilon_{22}. 
\eea
Noting
$q_{22}^2\ll\frac{(\delta{\tilde \omega}_{22}^2)^{\frac{1}{2}}}{H_0}$,
after matching these solutions,
at the leading order we obtain the coefficients
\bea
B_{+,+}
&\simeq &\frac{1}{(1-e^{-2\pi q_1})^{\frac{1}{2}}}
\Big(1
-\frac{m}{\sqrt{3}H_0}\Big)
e^{i\big(\frac{\pi}{4}
-\sqrt{
\frac{(\delta{\tilde\omega}_{22}^2)^{\frac{1}{2}}}{H_0}}\big)}
,
\nonumber\\
B_{+,-}
&\simeq &
\frac{e^{-\pi q_1}}
{(1-e^{-2\pi q_1})^{\frac{1}{2}}}
\Big(1
-\frac{m}{\sqrt{3}H_0}\Big)
e^{-i\big(\frac{\pi}{4}
-\sqrt{
\frac{(\delta{\tilde\omega}_{22}^2)^{\frac{1}{2}}}{H_0}}\big)}.
\eea

Then, at $t=t_{\ast}$ with ($|t_{\ast}|\ll |t_{1,+}|$),
where $t_{\ast}$ is chosen to be Eq. (\ref{mika2})
as for the non-planar high momentum modes,
the WKB solution are matched to
the de Sitter mode function
\bea
{\tilde H}_+^{(3)}&=&
\frac{C_{+,+}}{\sqrt{2k}}e^{i\frac{k}{H_0}(-3H_0t)^{\frac{1}{3}}}
\Big[\big(-3H_0 t\big)^{\frac{1}{3}}+\frac{iH_0}{k}\Big]
+
\frac{C_{+,-}}{\sqrt{2k}}e^{-i\frac{k}{H_0}(-3H_0t)^{\frac{1}{3}}}
\Big[\big(-3H_0 t\big)^{\frac{1}{3}}-\frac{iH_0}{k}\Big]
\nonumber\\
&\xrightarrow[t\to 0-]{}&
\frac{iH_0}{\sqrt{2k^3}}
\Big(
C_{+,+}-C_{+,-}
\Big).
\eea
For the high momentum modes,
\bea
C_{+,+}-C_{+,-}
\simeq
\frac{1
-\frac{m}{\sqrt{3}H_0}}
{(1-e^{-2\pi q_1})^{\frac{1}{2}}}
\Big\{
\big(1-e^{-\pi q_1}\big)
 \cos\big(\frac{\pi}{4}-\Psi_{+,2}\big)
+i\big(1+e^{-\pi q_1}\big)
 \sin\big(\frac{\pi}{4}-\Psi_{+,2}\big)
\Big\},
\eea
where
$\Psi_{+,1}:=
\sqrt{
\frac{(\delta{\tilde \omega}^2_{11})}
     {H_0(\delta{\tilde \omega}^2_{22})^{\frac{1}{2}}}}
+\Phi+\sqrt{\frac{k}{H_0}}$
and
$\Psi_{+,2}:=
\sqrt{
\frac{(\delta{\tilde \omega}^2_{22})^{\frac{1}{2}}}
     {H_0}}
+\Phi+\sqrt{\frac{k}{H_0}}$.
Thus, the power spectrum for the ${\tilde H}_{+}$
($=H_{+}$) mode is
given by
\bea
P_{h_+}
&\simeq&
P_{h_+}^{(0)}
\Big(1
-\frac{2m}{\sqrt{3}H_0}\Big)
\Big(
\coth(\pi q_1)
-\frac{\sin(2\Psi_{+,2})}{\sinh(\pi q_1)}
\Big),
\nonumber\\
\Psi_{+,2} &=&
\frac{\sqrt{\pi}\Gamma(\frac{1}{3})}{3\times 2^{\frac{1}{3}}\Gamma(\frac{5}{6})}\frac{k}{H_0}
+O(\sqrt{\frac{
k}{H_0}}).
\eea
Here we use the fact that $k_2 \simeq k$ in the case of planar modes.


\end{document}